\documentclass[]{spie}  

\usepackage{amsmath,amsfonts,amssymb}
\usepackage{graphicx}
\usepackage{multirow}
\usepackage[colorlinks=true, allcolors=blue]{hyperref}
\usepackage{xcolor}

\title{CCAT-prime: Optical and cryogenic design of the 850 GHz module for Prime-Cam}

\author[a]{Anthony I. Huber}
\author[b,c,d]{Scott C. Chapman}
\author[c]{Adrian K. Sinclair}
\author[e]{Locke D. Spencer}
\author[f]{Jason E. Austermann}
\author[g]{Steve K. Choi}
\author[c]{Jesslyn Devina}
\author[h]{Patricio A. Gallardo}
\author[d]{Doug Henke}
\author[g]{Zachary B. Huber}
\author[g]{Ben Keller}
\author[g,i]{Yaqiong Li}
\author[g]{Lawrence T. Lin}
\author[g,j]{Mike Niemack}
\author[k]{Kayla M. Rossi}
\author[i]{Eve M. Vavagiakis}
\author[f]{Jordan D. Wheeler}
\author[ ]{the CCAT-prime Collaboration}
\affil[a]{Dept. of Physics and Astronomy, University of Victoria, Victoria, Canada}
\affil[b]{Dept. of Physics and Atmospheric Science, Dalhousie University, Halifax, Canada}
\affil[c]{Dept. of Physics and Astronomy, University of British Columbia, Vancouver, Canada}
\affil[d]{NRC Herzberg Astronomy \& Astrophysics Research Centre, Victoria, Canada}
\affil[e]{Dept. of Physics and Astronomy, University of Lethbridge, Lethbridge, Canada}
\affil[f]{National Institute of Standards and Technology, Boulder, Colorado, USA}
\affil[g]{Dept. of Physics, Cornell University, Ithaca, USA}
\affil[h]{Kavli Institute for Cosmological Physics, University of Chicago, Chicago, USA}
\affil[i]{Kavli Institute at Cornell for Nanoscale Science, Cornell University, Ithaca, USA}
\affil[j]{Dept. of Astronomy, Cornell University, Ithaca, USA}
\affil[k]{Cornell Center for Astrophysics and Planetary Sciences, Cornell University, Ithaca, USA}

\authorinfo{Further author information: (Send correspondence to A.I.H.)\\A.I.H.: E-mail: ahuber@uvic.ca \\ S.C.C.: E-mail: Scott.Chapman@Dal.ca}

\begin{document} 
\maketitle

\begin{abstract}
Prime-Cam is a first-generation instrument for the Cerro Chajnantor Atacama Telescope-prime (CCAT-prime) Facility. The 850$~$GHz module for Prime-Cam will probe the highest frequency of all the instrument modules. We describe the parameter space of the 850$~$GHz optical system between the F$\lambda$ spacing, beam size, pixel sensitivity, and detector count. We present the optimization of an optical design for the 850$~$GHz instrument module for CCAT-prime. We further describe the development of the cryogenic RF chain design to accommodate $>$30 readout lines to read 41,400 kinetic inductance detectors (KIDs) within the cryogenic testbed.
\end{abstract}


\section{Introduction}
\label{sec:Introduction}

Far-infrared (FIR) astronomy plays a crucial role in our understanding of the Universe due to the fact that roughly half of all radiation incident from the Universe is observed on Earth within the FIR and submillimeter wavelength range\cite{Lutz}. In the decadal plans of both the National Aeronautics and Space Administration (NASA) and the European Space Agency (ESA), a series of questions were put forth\cite{CV,DS}. These include questions such as: What are the conditions for planet formation and the emergence of light? How does the Solar System work and how did we get here? How does the Universe work and what is it made of? FIR observations are well suited to address each of these questions as they provide powerful diagnostics
and are highly complementary to other types of observations\cite{Farrah}.

The CCAT-prime observatory utilizes a 6 m telescope designed for submillimeter to millimeter
wavelengths\cite{P2}, the Fred Young Submillimeter Telescope (FYST). The optical design of FYST consists of a novel off-axis crossed-Dragone design, allowing for high throughput and a large diffraction-limited field-of-view (FoV)\cite{Niemack:16}. Situated near the Cerro Chajnantor summit high in the Atacama Desert in Chile, CCAT-prime will be sited 5600 meters above sea level. With its high, dry site, CCAT-prime will be suitably positioned to negate much of the atmospheric effects, such as absorption due to atmospheric water vapor, that plague terrestrial infrared observatories\cite{P1}. The telescope optics are designed with high surface accuracy and low emissivity for superior surface brightness sensitivity in the submillimeter and millimeter atmospheric windows. Thus, particularly with the Prime-Cam instrument, CCAT-prime will provide unrivalled mapping speeds in these windows\cite{Stacey}, and be able to observe beyond the confusion limited depth of previous terrestrial and spaceborne observatories\cite{primecollaboration2021ccatprime}. 

The Prime-cam instrument is designed to incorporate independent modules, similar in design to the modules developed for ACTPol\cite{P2,Eve}, developed in collaboration with the Simons Observatory (SO). Similar cameras have been previously developed for astronomical observations\cite{P2}, but the unique capability of exploring the 850$~$GHz (350 $\mu$m) band makes this instrument crucial, as there are no
immediate proposals for another instrument with this capability\cite{Scott}. The designs of the 850$~$GHz instrument module require opto-mechanical structures, filtering, silicon lenses, readout electronics, control and data processing software, and cryogenics.

This report outlines the design of the 850$~$GHz module and its vital role in Prime-Cam. The structure of this report is as follows: in Sec.$~$\ref{sec:Overview} an overview of Prime-Cam will be presented, highlighting the relevance of the 850$~$GHz module; Sec.$~$\ref{sec:OpticalDesign} will describe the considerations and parameter space of the design, concluding with the current optical design of the 850$~$GHz module; the preliminary design work for the radio frequency (RF) chain for the detector arrays are presented in Sec.$~$\ref{sec:RFChainDesign}; the conclusion, which highlights future work, is given in Sec.$~$\ref{sec:Conclusion}.

\section{Prime-Cam Overview}
\label{sec:Overview}

The CCAT-prime observatory has been designed to answer fundamental astrophysical questions ranging in scope from Big Bang cosmology and the large-scale structure of the Universe down to the formation of stars and planetary systems in our own Galaxy. 
Such studies require high-sensitivity polarimetric, photometric, and spectroscopic mapping at several frequencies across the relevant band of the electromagnetic spectrum\cite{Stacey}.
The Prime-Cam instrument has been designed to explore several science goals\cite{primecollaboration2021ccatprime}.
First, through Prime-Cam the evolution of dusty star formation into the epoch of galaxy assembly ($>$ 10 billion years ago) will be traced through multi-frequency photometric measurements.
This work will also lead to improvements to the constraints on primordial gravitational waves and inflationary models by characterizing CMB foreground dust polarization. Likewise, through intensity mapping Prime-Cam will reveal the formation, growth, and three-dimensional large scale clustering properties of the first star-forming galaxies within the first billion years after the Big Bang\cite{Kovetz}. Lastly, Prime-Cam will also explore the physical properties and spatial distribution of galaxy clusters via the Sunyaev-Zeldovich effects on the CMB, constraining fundamental physics such as dark energy, neutrino masses, and active galactic nuclei-star formation feedback mechanisms\cite{Mittal}.

\begin{figure}[t]
    \centering
    \includegraphics[width=0.725\linewidth]{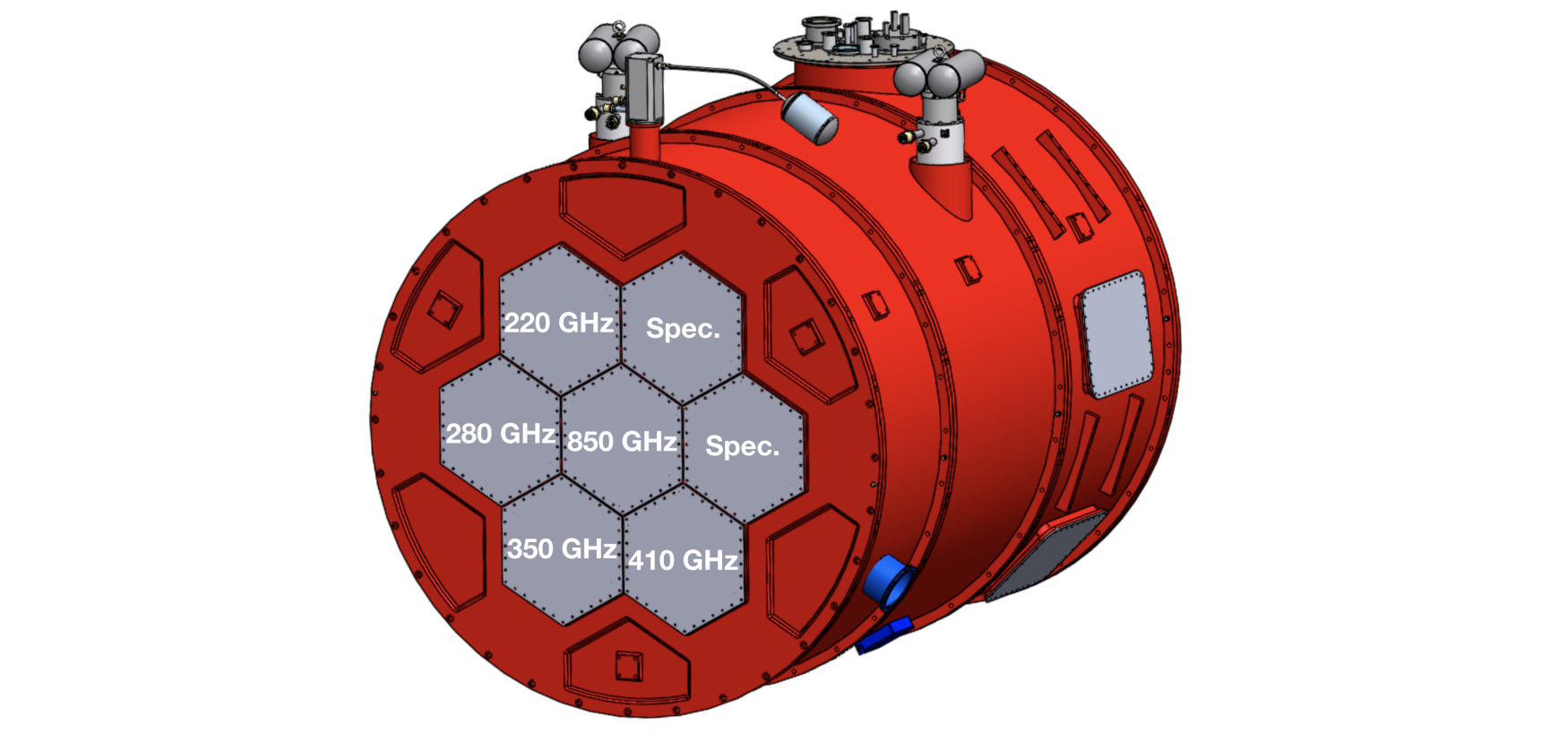}%
    \caption{Rendering of Prime-Cam highlighting the modular nature of the instrument. The baseline design for each module incorporates three silicon lenses, KID arrays cooled to $\sim 100$$~$mK, and a complex readout system. Image taken from \citenum{Choi:2020a}.}%
    \label{fig:PCam}%
\end{figure}

Of the 8$^\circ$ diameter FoV from FYST, roughly 4.9$^\circ$ is filled by Prime-Cam.
This is achieved by incorporating a series of distinct instrument modules, as shown in Fig.$~$\ref{fig:PCam}, with six instrument modules arranged around a seventh central module. Each of these modules fills a 1.3$^\circ$ diameter FoV, with 1.8$^\circ$ separation\cite{P2,primecollaboration2021ccatprime}.
The baseline designs for Prime-Cam incorporate two imaging spectrometer modules utilizing Fabry-Perot Interferometers for line intensity mapping from 210 to 420$~$GHz\cite{Nikola}, and five polarization-sensitive modules at frequencies of 220$~$GHz, 280$~$GHz, 350$~$GHz, 410$~$GHz, and 850$~$GHz\cite{Choi:2020a}.
Each of these modules will incorporate kinetic inductance detector (KID) arrays due to ease of fabrication and readout while achieving the fundamental background limits required for Prime-Cam\cite{Hubmayr}.
The overall configuration of Prime-Cam is shown in Fig.$~$\ref{fig:PCam}.

The 850$~$GHz module is of critical importance to extract astrophysical properties such as infrared luminosities and dust-obscured star formation rates in dusty star forming galaxies\cite{primecollaboration2021ccatprime}.
This module may therefore be used to constrain the peak of galactic dust spectral energy distributions, enabling precision estimates of source luminosity in the context of the evolution of these galaxies\cite{Elbaz}.
850$~$GHz surveys will explore fainter sources than Herschel, and greater volumes than ALMA, increasing the likelihood of revealing rare sources\cite{Stacey,primecollaboration2021ccatprime}.
These requirements demand an instrument module with a large FoV and exceptional spatial resolution.
Given the vital science achievable with this instrument, the 850$~$GHz module will occupy the privileged central position in Prime-Cam to aid in preserving the image quality.
However, as the highest frequency module there are additional design considerations which must be made beyond the other instrument modules.

\section{Optical Design}
\label{sec:OpticalDesign}

This section will first highlight an attempt to incorporate the SO instrument module design into the 850$~$GHz instrument module of Prime-Cam as presented in \citenum{Eve}, and then detail the first full optical design of the 850$~$GHz instrument module including a discussion of the parameter space considered in this work.
The optical configurations presented are extensions of the three-lens cameras proposed for the SO\cite{SO}. While they do share basic architecture, the designs across all the instrument modules for Prime-Cam have been modified to accommodate the shorter wavelengths and stringent requirements for the Strehl ratio across the FoV. However, each of the designs benefit indirectly from the thoroughly studied systematics of these designs\cite{Pato,Gudmundsson:21}.
The design work utilized the Zemax OpticStudio software\cite{Zemax}.

\subsection{Simons Observatory Instrument Module Design}
\label{subsec:PreliminaryDesign}

The science goals for the 850$~$GHz module demand high resolution with minimal degradation to mapping speed. These demands create a tradeoff between F$\lambda$ spacing, beam size, pixel sensitivity, and detector count.
To an extent, both the efficiency and sensitivity increase with F$\lambda$, while resolution and mapping speed decrease.
As a baseline, each Prime-Cam instrument module should be designed to: 1)$~$have a 1.3$^\circ$ diameter FoV, 2)$~$have high image quality, which is assessed based on the Strehl ratio across the FoV with a goal of $>$0.8, and 3)$~$maximize the mapping speed.  
The baseline detector count for the 850$~$GHz module is roughly 20,000 KIDs, while a detector count as high as 50,000 is attainable.
However, additional detectors require considerations to the thermal loading of the instrument, which will be discussed in Sec.$~$\ref{sec:RFChainDesign}.

\begin{figure}[t]
    \centering
    \includegraphics[width=\linewidth]{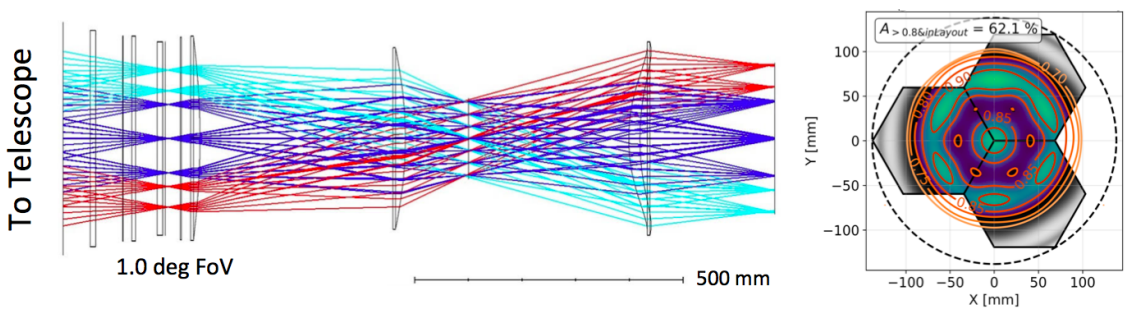}%
    \caption{Left:$~$Optical design of the 850$~$GHz instrument module utilizing the base SO design. Right:$~$Simulation of the Strehl ratio across the FoV for the SO instrument module design for 850$~$GHz. The initial three-lens design resulted in low image quality and reduced FoV (1$^\circ$) compared to other instrument modules. Image taken from \citenum{Eve}.}%
    \label{fig:prelim}%
\end{figure}

Each of the Prime-Cam instrument modules are designed to provide diffraction-limited image quality across the wide FoV provided from FYST.
The initial design of the 850$~$GHz module attempted to match the designs of the SO instrument modules in Prime-Cam.
As can be seen in Fig.$~$\ref{fig:prelim}, one can incorporate the SO design into the Prime-Cam instrument modules to include three cryogenically cooled lenses, a collimated beam at a 1$~$K Lyot stop, and wide FoV focused onto a densely populated detector array.
While for the lower frequency instrument modules this configuration proves effective, for the 850$~$GHz module this design results in a significantly reduced FoV (1$^\circ$), as shown in the right panel of Fig.$~$\ref{fig:prelim}, with only 62.1$\%$ of the FoV within the acceptable limit for the Strehl ratio.
This motivates the need for additional study beyond the work done for the SO instrument module design to develop a system capable of achieving the Prime-Cam requirements for the 850$~$GHz module in terms of the instrument sensitivity, mapping speed, image quality, and FoV.
The preliminary SO design serves as the baseline for comparison with the 850$~$GHz instrument module.

\subsection{850 GHz Instrument Module Design}
\label{subsec:CurrentDesign}

\subsubsection{Parameter Space}
From the outset of this study there were constraints to be considered in the design of the system.
One constraint in the 850$~$GHz optical design is the choice of silicon for the lens material.
The choice of cryogenic silicon was due to its excellent optical performance at the Prime-Cam frequencies and thermal conductance at cryogenic temperatures.
Since silicon is unparalleled in terms of its optical properties and thermal conductivity, no other lens materials were explored.
Note this prevents configurations of lenses with mixed indexes of refraction, such as the Cooke triplet. 

A second constraint arises from the process of applying the anti-reflection (AR) coating to the lenses and the physical length of the Prime-Cam instrument itself.
The application of the AR coating to the lens requires that the sagitta of the lens be less that 14 mm, else the manufacture time and difficulty increase and some losses in the effectiveness of the AR coating can occur\cite{Jeff}.
Since the index of refraction may be, to first order, taken as proportional to the frequency, the 850$~$GHz module contends with the most extreme index of refraction in Prime-Cam.
From the outset of this study it became apparent that there are two issues: 1) the available instrument module space in Prime-Cam prevents thinner lenses to be used in the 850$~$GHz module, and 2) each additional lens reduces the overall throughput of the system by 4\%, setting a constraint on the number of lenses possible in the system.
Thus, this study explored a range of systems using three to five lenses, resulting in the FoV ranging from 0.7$^\circ$-1.3$^\circ$, depending on the number of lenses used.

Throughout the design work for the 850$~$GHz module, several variations were explored to cover the parameter space of the instrument. Of particular interest were the length of the optics, number of lenses, and field curvature at the image plane. The first of these, length, was simple to constrain. 
It was clearly shown that any length beyond the $\sim$1000 mm limit of Prime-Cam resulted in both increased Strehl and FoV.
When the system was optimized without constraining length for a four-lens system it was found that the ideal length was $\sim$1200 mm, which exceeds the maximum limit of the Prime-Cam design by $\sim$200 mm and is well beyond what can be conceivably accommodated.
At shorter lengths it is clear that the Strehl ratio and FoV become, to an extent, competing variables.
With a reduction in length, faster - and therefore thicker - lenses are required to the point that the limits imposed by the AR coating are exceeded.

In order to reduce the lens thicknesses and compensate for the reduced optical length of the 850$~$GHz module, a fifth lens was added to the designs.
With a fifth lens included, the designs were able to achieve a 1.3$^\circ$ FoV.
However, this design was unable to completely reduce the thickness of all the lenses to avoid fabrication issues for the AR coating.
This confirms that having one lens exceeding the sagitta limit for the AR coating is unavoidable.
Additionally, the inclusion of a fifth lens reduces the throughput of the instrument module by 4$\%$, which results in a significant reduction in the mapping speed of the module\cite{Scott2022}.
It is clear that unless the number of the detectors is pushed beyond 45,000, there is no real advantage to the inclusion of a fifth lens.

\begin{figure}[t]
\centering
\includegraphics[width=0.8\linewidth,height=10cm]{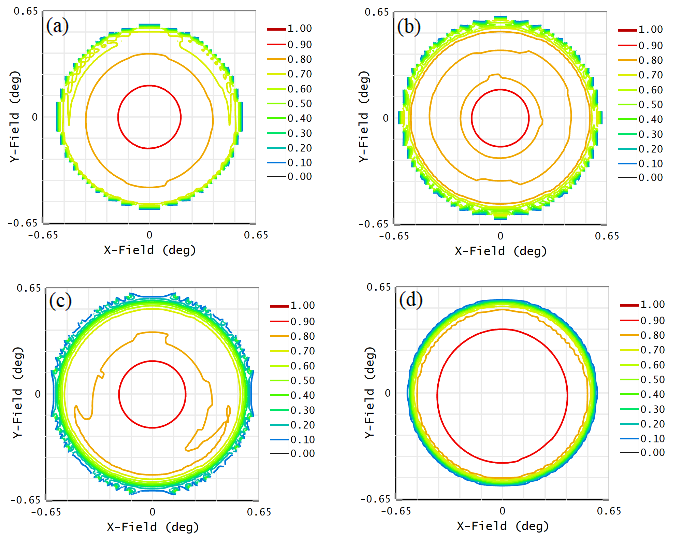}
    \caption{Simulations of the Strehl ratio across variations in the field curvature represented by the maximum allowable tilt: (a) 2$^\circ$, (b) 4$^\circ$, (c) 5$^\circ$, and (d) 8$^\circ$. Note that the Stehl ratio clearly improves as the field curvature increases, but at the cost of a potential reduction in detector sensitivity and increased cross-polarization.}%
    \label{fig:curve}%

\end{figure}

The final variable explored in this work is the degree of allowable field curvature at the image plane of the system.
This idea is being explored as an alternate means of compensating for the reduced optical length of the instrument module, which can be represented to first order as a tilt effect introduced at the detector array with the degree of tilt increasing radially across the array.
As shown in Fig.$~$\ref{fig:curve}, increasing the maximum tilt, or angle of incidence, in the design results in a wider FoV and improvements to the Strehl ratio.
Even so, the inclusion of tilt beyond 2$^\circ$ may impact the effectiveness of the detector feedhorns, and therefore the sensitivity of the instrument.
The reduction in sensitivity can be partially mitigated by having a system with smaller F$\lambda$ spacing, corresponding to a higher detector count over a given FoV.
While increasing the detector count results in more spillover losses\cite{Griffin}, more detectors results in a net increase in mapping speed in the ranges considered in this work (see \citenum{Scott2022} for a full analysis).
However, high detector counts present challenges to readout design and thermal budgeting, as will be discussed in Sec.$~$\ref{sec:RFChainDesign}.
The overall impact of field curvature on factors such as sensitivity, efficiency, and cross-polarization are still being explored.
Nevertheless, as the largest impact of field curvature lies on the edge of the FoV where there are fewer detectors a baseline design allowing up to a maximum of 4$^\circ$ tilt has been used for this work under the assumption the reduction in sensitivity due to tilt is minimal compared to the gain in the Strehl ratio across the FoV.

Note that all of the issues presented in the parameter space are exacerbated at higher frequencies, an important consideration for proposed high-frequency instrument modules such as the 1500$~$GHz module\cite{Stacey}.

\subsubsection{Optical Design}

\begin{figure}[t]
    \centering
    \includegraphics[width=0.9\linewidth]{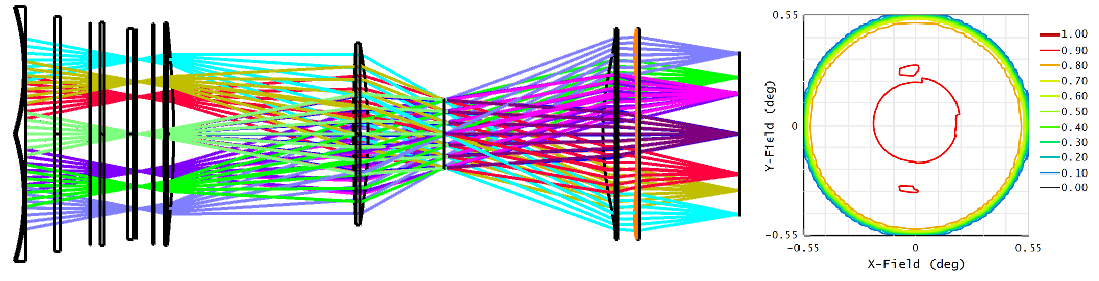}%
    \caption{Current design of the 850$~$GHz module for Prime-Cam and simulation of the Strehl ratio across the FoV. The four-lens system results in high image quality across a 1.1$^\circ$ diameter FoV.}%
    \label{fig:current}%
\end{figure}

The current optical design of the 850$~$GHz module for Prime-Cam is presented in Fig.$~$\ref{fig:current}.
To maximize the mapping speed of the instrument module, a four-lens design was selected over a five-lens design.
Furthermore, the FoV was reduced from 1.2$^\circ$ to 1.1$^\circ$ to achieve the highest possible Strehl ratio over the largest area of the FoV possible.
The result is a system with a Strehl ratio $>0.8$ across nearly the entire FoV.
This is accomplished in part by allowing a maximum angle of incidence of 4$^\circ$.
The design includes 41,400 KIDs across three silicon feedhorn arrays, which corresponds to a pixel pitch of 1.42 mm and an F$\lambda$ spacing of 1.62.
Since the value of F$\lambda$ is reduced from the baseline value of 2.0, the system will be less susceptible to the effects of field curvature.
Thus the design is a robust solution to the complex parameter space and constraints described above.
The final variable, which will be described in detail below, is the RF chain development which will constrain the thermal budget of the system and, therefore, the maximum number of detectors available in the 850$~$GHz module.

\section{RF Chain Design}
\label{sec:RFChainDesign}
\begin{figure}[b]
    \centering
    \includegraphics[width=0.9\linewidth,height=4.75cm]{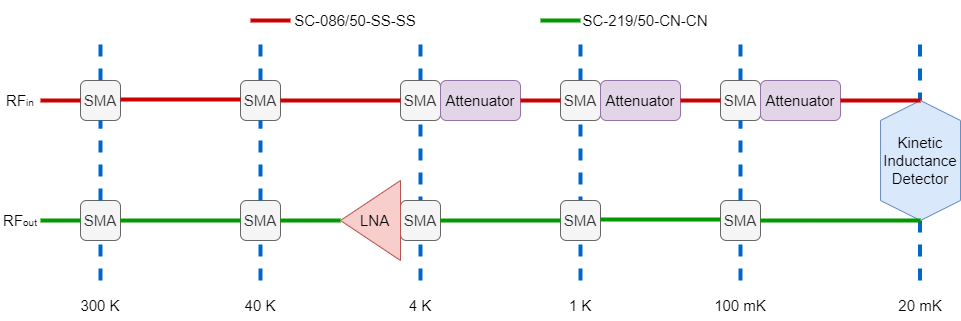}%
    \caption{Schematic of the RF chain design for the 850$~$GHz module as it is to be incorporated into the cryogenic testbed. The components will be thermally coupled to the relevant stages to prevent thermal parasitics from reaching the KID arrays.}%
    \label{fig:RFSchematic}%
\end{figure}

The cryogenic testbed for the 850$~$GHz module uses a commercial Bluefors LD250 dilution refrigerator\cite{Bluefors}.
The system is designed to provide a large volume for experiments with several thermally isolated stages with the base stage reaching a temperature below 10$~$mK in less than 24 hours.
Each LD250 unit contains numerous access ports, which are vital for testing KID arrays with large RF demands.
The cryostat is designed to thermalize the semi-rigid coaxial cables between each of the stages, allowing for a reduction in thermal noise.
As can be seen in Fig.$~$\ref{fig:RFSchematic}, the design for the RF chain as installed in the LD250 system will also incorporate cryogenic attenuators and a cryogenic low noise amplifier (LNA).
For the simulations that follow, the system assumes the use of SC-086/50-SS-SS coaxial cables feeding on the input side of the chain, and SC-219/50-CN-CN on the output side\cite{coax}.
However, in Prime-Cam the 850$~$GHz module will incorporate flexible stripline based transmission lines\cite{Neric}.
The readout hardware and software external to the cryostat are being developed to double the available bandwidth and are presented in a companion paper\cite{Adrian}.

\subsection{Thermal Budget}
\label{subsec:thermal}

For the cryostat to run as effectively as possible, thermal loading must be kept to a minimum.
As the LD250 dilution refrigerator is a commercial unit, care has already been taken in the design to reduce thermal loading from radiative transfer and thermal conductance.
However, the inclusion of the RF chain introduces new thermal loading via thermal conductance across the coaxial cables and dissipation from the cryogenic attenuators and LNAs.
With more detectors, more cables, attenuators, and LNAs are required.
This is the motivation for doubling the readout bandwidth to halve the required hardware for a given detector array\cite{Adrian}.

From the current optical design detailed above, the 850$~$GHz module is proposed to include 41,400 KIDs across three wafer arrays, the details of which are given in a companion paper\cite{Scott2022}.
Assuming the readout bandwidth is doubled, the 850$~$GHz module will require 9 Radio Frequency System-on-Chip (RFSoC) boards\cite{RFSoC}, 36 cryogenic LNAs, 108 cryoattenuators, and 72 transmission lines.
A rough thermal budget for implementation in Prime-Cam is used based on a preliminary analysis of the instrument as these constraints are tighter than those of the LD250 testbed.
For the cryoattenuators, the thermal budget and subsequent analyses assume attenuation values of 23$~$dB, 10$~$dB, and 3$~$dB for the 4$~$K, 1$~$K, and 100$~$mK stages respectively.
Additionally, the maximum dissipation per LNA was set to 8 mW.
A comparison of the thermal loading introduced by the RF chain with the preliminary thermal budget is shown in Tab.$~$\ref{tab:Budget}.
While the thermal loading for all components are within the constraints of the thermal budget, the results emphasize the importance of doubling the bandwidth to halve the requisite number of LNAs.
Without this key development in the RF readout, the detector count would need to be reduced to roughly 30,000 KIDS so that dissipation from the LNAs would fall within the thermal budget for the 4$~$K stage.

\begin{table}[h]
\caption{Comparison of the predicted thermal loading for RF chain components in the 850$~$GHz module compared with the overall thermal budget at each stage based on preliminary values for Prime-Cam.} 
\label{tab:Budget}
    \begin{center}
        \begin{tabular}{|l|c|c|}
            \hline
            \rule[-1ex]{0pt}{3.5ex} Stage/Component & Thermal Loading & Thermal Budget \\ \hline
            \rule[-1ex]{0pt}{3.5ex} 40$~$K Coax & 0.278 W & 12.57 W  \\\hline
            \rule[-1ex]{0pt}{3.5ex} 4$~$K Coax & 7.68 mW & \multirow{3}*{457.14 mW}  \\\cline{1-2}
            \rule[-1ex]{0pt}{3.5ex} 4$~$K Attenuator & 0.008 mW & ~  \\\cline{1-2}
            \rule[-1ex]{0pt}{3.5ex} 4$~$K LNA & 288 mW & ~  \\\hline
            \rule[-1ex]{0pt}{3.5ex} 1$~$K Coax & 0.0401 mW & \multirow{2}*{1.71 mW}   \\\cline{1-2}
            \rule[-1ex]{0pt}{3.5ex} 1$~$K Attenuator & 0.0002 mW & ~   \\\hline
            \rule[-1ex]{0pt}{3.5ex} 100$~$mK Coax & 1.93 {\textmu}W & \multirow{2}*{45.71 {\textmu}W}   \\\cline{1-2}
            \rule[-1ex]{0pt}{3.5ex} 100$~$mK Attenuator & 0.0206 {\textmu}W & ~   \\\hline
            \rule[-1ex]{0pt}{3.5ex} 20$~$mK Coax & 0.0328 {\textmu}W & $\sim$14 {\textmu}W \\ \hline
        \end{tabular}
    \end{center}
\end{table}

\subsection{Scattering Parameters}
\label{subsec:sParam}

Scattering parameters, or S-parameters, are a set of coefficients used to describe the response of an electrical network, particularly for those operating at RF and microwave frequencies.
S-parameters are used to relate the magnitude and phase of voltage waves incident to and reflected from network ports\cite{RFBasics}.
However, S-parameters change with measurement frequency, meaning that characterizing a system requires an analysis of each component across the band of interest; some components may be calculated directly using network analysis, while others require measurement using a vector network analyzer\cite{Pozar}.
These results are then cascaded through the system to determine the overall S-parameters, which for a two-port system such as the RF chain design in the 850$~$GHz module allows for the overall attenuation in a system to be quickly and accurately simulated.
An additional advantage of the S-parameter based model is for debugging RF link breaks which would show up as anomalously large reflections in S$_{11}$ or S$_{22}$.
Factors include the LNAs, cryoattenuators, and KIDs, but for accuracy should include losses in the coaxial cables.
Values for the attenuation vary between different types and manufacturers, so it is vital to ensure the correct parameters are used for simulations.
Furthermore, adding the measured S-parameters as each component is acquired will help strengthen the model's predictive abilities.

The 850$~$GHz module is designed to introduce as much attenuation as possible (at least 30 dB) before reaching the detector arrays, and as little attenuation as possible onward to the output of the cryostat.
Consequently, a series of simulations were performed\cite{2022Arsenovic} first using different types of coaxial cables to determine which would be ideal for the system.
From this analysis the SC-086/50-SS-SS coaxial cables were selected for the input side of the chain and SC-219/50-CN-CN for the output side as shown in Fig.$~$\ref{fig:RFSchematic}.
Additionally, the simulations included the S-parameters for two KIDs with resonant frequencies of 1$~$GHz and 1.001$~$GHz, with total quality factors of 4,000 and 3,000 respectively and a coupling quality factor of 10,000 for both.
The S-parameters were calculated assuming a symmetric system via the method presented in \citenum{Thomas_2015}.
If provided with the de-embedded S-parameters of the 850$~$GHz array, they can be fully simulated within this model.

\begin{figure}%
    \centering
    \includegraphics[width=0.5\linewidth]{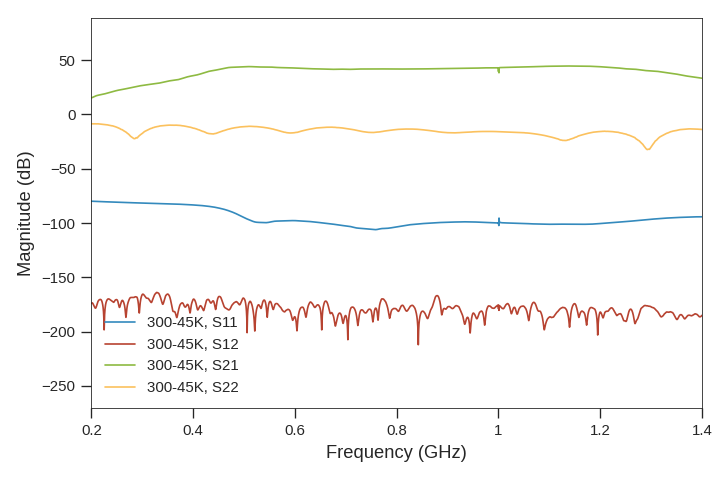}%
    \includegraphics[width=0.5\linewidth]{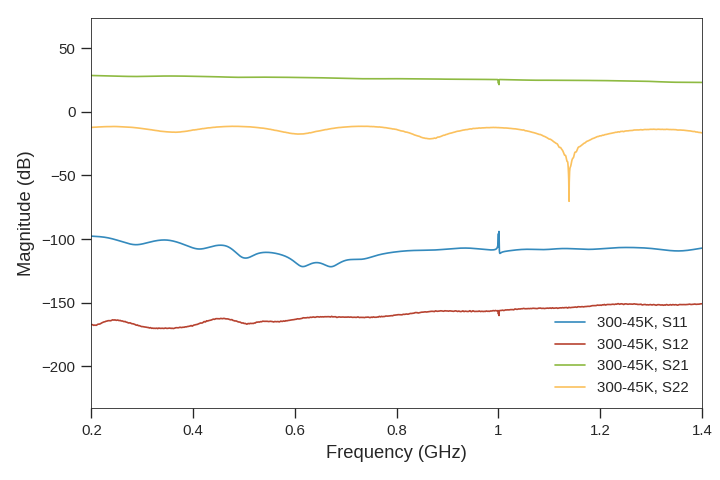}%
    \caption{Simulated S-parameters through the 850$~$GHz cryogenic testbed, including detector resonances at both 1$~$GHz and 1.001$~$GHz. The simulations considered attenuation in cables, as well as cryogenic attenuators and low noise amplifiers (LNAs), including a room-temperature LNA external to the cryostat. The left panel depicts simulations using a high-gain LNA, while the right panel depicts a low-dissipation LNA.}%
    \label{fig:gain}%
\end{figure}

Lastly, a series of LNAs were explored to examine the attenuation in the RF chain against power dissipation.
S-parameters for the LNA were taken from the manufacturers for these simulations.
The results, shown in Fig.$~$\ref{fig:gain}, highlight these two parameters.
Note that the expected band for the readout is approximately 0.3$~$GHz to 1.3$~$GHz.
The left panel of Fig.$~\ref{fig:gain}$ depicts a system utilizing a high-gain system operating at the maximum allowable dissipation of 8 mW\cite{Jiang2014}, while the left panel depicts a system with a dissipation of 5.2 mW\cite{CryoElecLNA}; the former reduces but does not eliminate the requirements for signal amplification outside the cryostat, while the latter drastically reduces thermal loading.
An additional room-temperature LNA\cite{zkl_1r5} was included prior to being connected to the RFSoC, the justification for which is presented below.
These simulations must be compared against in situ measurements before the system can be finalized.

\subsection{Cascaded Noise Temperature}
\label{subsec:noise}

The final step in characterizing the RF chain is an analysis of the noise inputs.
The cascaded noise temperature from the output of the detector to the output port of the cryostat (or onward) can be calculated using the Friis formula,
\begin{equation}
    T_{cas} = T_{det} + T_{cable 1} + \frac{T_{cable 2}}{G_{cable 1}} + \frac{T_{lna}}{G_{cable 1} G_{cable 2}}+...,
\end{equation}
\noindent where $G_x = |S_{21}|^2_x$ is the gain of the preceding stage $x$ calculated from its forward transmission scattering parameter. $T_{cable 1}$ is the effective noise temperature of the first cable calculated from its loss L and physical temperature as $T_{cable 1}= T_{phy}(L-1)$.

\begin{figure}[t]
    \centering
    \includegraphics[width=0.8\linewidth,height=10cm]{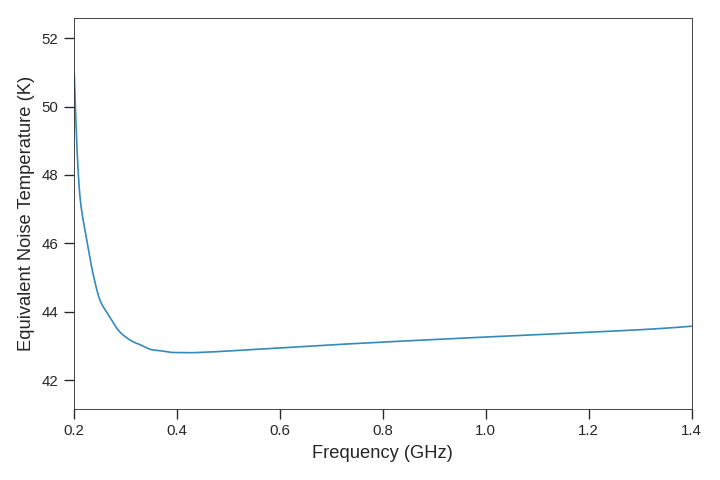}%
    \caption{Simulated equivalent noise temperature across the detector array readout bandwidth from the cryogenically cooled KID array to the output of the RT amplifier\cite{zkl_1r5}. The simulation assumes lossy cables and an equivalent output noise temperature of 40$~$K when on resonance based on values from the BLAST-TNG 850$~$GHz array.}%
    \label{fig:noise}%
\end{figure}

In the calculation of the equivalent noise temperature of the cryogenic testbed for the 850$~$GHz module, an equivalent output noise temperature of 40$~$K was used for the KID arrays based on calculations from \citenum{Sipola2019} using values measured for the BLAST-TNG 850$~$GHz array\cite{AdrianThesis}.
A derivation of this value is given in \citenum{Adrian}, and the equivalent noise temperature as calculated across the band of the RF chain in the cryostat with an additional room-temperature amplifier is shown in Fig.$~$\ref{fig:noise}.
With a baseline detector equivalent noise temperature of 40$~$K, the added noise from the output of the detector to the output of the room-temperature amplifier is less than 10\%.

The equivalent noise temperature of the RFSoCs analog to digital converters can be calculated using the worst-case noise spectral density (NSD) value of -141 dBFS/Hz and full scale input power, $P_{FS}$ = 2 dBm, from \citenum{DS926}. Using these numbers we get,
\begin{equation}
    T_{adc} = 10^{(P_{FS}+NSD)/10}/1000/k \approx 10^6 \ \textrm{K}
\end{equation}
where $k$ is Boltzmann's constant.
$T_{adc}$ can then be used to determine the total cascaded noise temperature of the total system.

From the cryostat, the signal will be amplified by a room-temperature RF amplifier, such as the ZKL-1R5+ by Mini-Circuits\cite{zkl_1r5}, providing $\sim$ 40 dB of gain and a noise temperature of $\sim$ 300 K.
Directly after the amplifier and before the RFSoC ADC, a programmable attenuator with a range of 30$~$dB will be used to optimize the input power so as to maximize the dynamic range but not saturate the ADC. 
We can calculate the optimal gain from the full scale input power of the ADC (2 dBm) and the total frequency comb power.
A typical detector requires a bias power of -90 dBm\cite{AdrianThesis} and a 1000 tone waveform would give -60 dBm total power at the input to the LNA. The optimal gain can be calculated as,
\begin{equation}
    G_{opt} = P_{ADCmax}/P_{comb} \approx 62 \ \textrm{dB}.
\end{equation}
This exceeds the gain of both LNAs in this study and motivates the use of additional gain which can be applied at room temperature. The combination of the ZKL-1R5+ RF amplifier and the following programmable attenuator allows for tuning to the optimal gain.

The total noise temperature including the ADC is represented as
\begin{equation}
    T_{tot}= T_{cas} + T_{adc}/G_{opt},
\end{equation}
where $G_{opt}$ is the optimal gain of the cascade from the input of the LNA to the input of the ADC.
The ratio of the simulated detector noise to the total noise is then given as
\begin{equation}
    \frac{T_{det}}{T_{tot}} = \frac{T_{det}}{T_{cas} + T_{adc}/G_{opt}} \approx 0.92.
\end{equation}
Thus our system would add 8\% to the total noise in the worst case with the optimal gain. If the 850$~$GHz detector arrays can be biased at higher powers this number will come down to less than 8\%.

\section{Conclusion}
\label{sec:Conclusion}

This paper detailed the parameter space considered for both the optical design of the 850$~$GHz module of the Prime-Cam instrument and the accompanying RF chain development for the cryogenic testbed.
Given the constraints outlined above, an optical design has been presented as a new baseline for the 850$~$GHz module with a predicted mapping speed $\sim$80\% greater\cite{Scott2022} than the previous baseline despite not achieving a 1.3$^\circ$ FoV.
The new baseline results in a design which will incorporate 41,400 KIDs across three silicon arrays.
With this detector count the system will have an F$\lambda$ spacing of $\sim$1.62, meaning the system will be less sensitive to field curvature at the image plane.
Once these parameters are finalized, the optical design will undergo a preliminary design review, with fabrication scheduled to begin late 2022 to meet the 2025 deployment goal.

Analysis of the RF chain parameters resulted in performance of the system to be compared against the cryogenic testbed once installation is complete.
The S-parameters provide a baseline comparison of potential RF designs, and additionally provide a model for debugging the RF chain for link breaks, poor connections, and faulty components.
Furthermore, the cascaded noise temperature highlights where the RF chain can be improved to provide a fully optimized testbed for characterizing the 850$~$GHz module.
Work on characterizing the RF chain will commence once the LD250 dilution refrigerator is installed, with work characterizing sample KIDs and readout software to follow.

\acknowledgments
The CCAT-prime project, FYST and Prime-Cam instrument have been supported by generous contributions from the Fred M. Young, Jr. Charitable Trust, Cornell University, and the Canada Foundation for Innovation and the Provinces of Ontario, Alberta, and British Columbia. The construction of the FYST telescope was supported by the Gro{\ss}ger{\"a}te-Programm of the German Science Foundation (Deutsche Forschungsgemeinschaft, DFG) under grant INST 216/733-1 FUGG, as well as funding from Universit{\"a}t zu K{\"o}ln, Universit{\"a}t Bonn and the Max Planck Institut f{\"u}r Astrophysik, Garching.


\bibliography{main} 
\bibliographystyle{spiebib} 

\end{document}